\documentclass[twocolumn,superscriptaddress,prb,final,showpacs,preprintnumbers,amsmath,amssymb,floatfix]{revtex4}
\usepackage{graphicx}
\setkeys{Gin}{width=3.4in,keepaspectratio}
\usepackage{dcolumn}
\usepackage{bm}

\begin{document}
  \newcommand*{\NCCO}[0]{Nd$_{2-x}$Ce$_{x}$CuO$_{4\pm\delta}$}
  \newcommand*{\LSCO}[0]{La$_{2-x}$Sr$_x$CuO$_4 $}
  \newcommand*{\LNSCO}[0]{La$_{1.6-x}$Nd$_{0.4}$Sr$_x$CuO$_4$}
  \newcommand*{\YBCO}[0]{YBa$_2$Cu$_3$O$_{6+\delta}$}
  \newcommand*{\NCO}[0]{Nd$_2$CuO$_{4\pm\delta}$}
  \newcommand*{\LCZMO}[0]{La$_2$Cu$_{1-z}$(Zn,Mg)$_z$O$_4$}
  \newcommand*{\NDO}[0]{(Nd,Ce)$_2$O$_3$}
  \newcommand*{\CuIon}[0]{Cu$^{\text{2+}}$}
  \newcommand*{\Neel}[0]{N\'{e}el}
  \newcommand*{\degC}[0]{\ensuremath{^\circ}C}
  \newcommand\degrees[1]{\ensuremath{#1^\circ}}
  \newcommand*{\etal}[0]{\textit{et~al.}}
  \newcommand*{\xiab}[0]{$\xi_\text{ab}$}
  \newcommand*{\xic}[0]{$\xi_\text{c}$}
  \newcommand*{\Tc}[0]{$T_\text{c}$}

\graphicspath{{StructureFigures/}{ReductionFigures/}{SSRLFigures/}{ChalkRiverFigures/}{10NSC/}
  {NatureComment/}{DataFromDai/}{TEM_Photos/}}

\title{Phase Decomposition and Chemical Inhomogeneity in \NCCO}

\author{ P.K. Mang }
\affiliation{Department of Applied Physics, Stanford University,
Stanford, California 94305}

\author{ S. Larochelle }
\altaffiliation {Present Address: Department of Physics,
University of Toronto, Toronto, Ontario M5S 1A7, Canada}
\affiliation{Department of Physics, Stanford University, Stanford,
California 94305}

\author{A. Mehta}
\affiliation{Stanford Synchrotron Radiation Laboratory, Stanford
University, Stanford, California 94309}

\author{O.P. Vajk}
\altaffiliation {Present Address: NIST Center for Neutron
Research, National Institute for Standards and Technology,
Gaithersburg, Maryland 20899}
\affiliation{Department of Physics,
Stanford University, Stanford, California 94305}

\author{A.S. Erickson}
\affiliation{Department of Applied Physics, Stanford
University, Stanford, California 94305}

\author{L. Lu}
\affiliation{Department of Applied Physics, Stanford University,
Stanford, California 94305}

\author{W.J.L. Buyers}
\affiliation{National Research Council, Neutron Program for
Materials Research, Chalk River Laboratories, Chalk River, Ontario
K0J 1J0 Canada}

\author{A.F. Marshall}
\affiliation{T.H. Geballe
Laboratory for Advanced Materials, Stanford University, Stanford,
California 94305}

\author{K. Prokes}
\affiliation{Hahn-Meitner
Institute, Glienicker Str. 100, Berlin D-14109, Germany}

\author{M. Greven}
\affiliation{Department of Applied Physics,
Stanford University, Stanford, California 94305}
\affiliation{Stanford Synchrotron Radiation Laboratory, Stanford
University, Stanford, California 94309}

\date{\today}

\begin{abstract}
Extensive X-ray and neutron scattering experiments and additional
transmission electron microscopy results reveal the partial
decomposition of \NCCO~(NCCO) in a low-oxygen-fugacity environment
such as that typically realized during the annealing process
required to create a superconducting state. Unlike a typical
situation in which a disordered secondary phase results in diffuse
powder scattering, a serendipitous match between the in-plane
lattice constant of NCCO and the lattice constant of one of the
decomposition products, \NDO, causes the secondary phase to form
an oriented, quasi-two-dimensional epitaxial structure.
Consequently, diffraction peaks from the secondary phase appear at
rational positions ($H,K,0$) in the reciprocal space of NCCO.
Additionally, because of neodymium paramagnetism, the application
of a magnetic field increases the low-temperature intensity
observed at these positions via neutron scattering. Such effects
may mimic the formation of a structural superlattice or the
strengthening of antiferromagnetic order of NCCO, but the
intrinsic mechanism may be identified through careful and
systematic experimentation.  For typical reduction conditions, the
\NDO~volume fraction is $\sim 1\%$, and the secondary-phase layers
exhibit long-range order parallel to the NCCO CuO$_2$ sheets and
are $50-100$ \AA~thick.  The presence of the secondary phase
should also be taken into account in the analysis of other
experiments on NCCO, such as transport measurements.
\end{abstract}


\maketitle

\section{Introduction}
Since its discovery\cite{NCCO:Discovery} in 1989, the
electron-doped superconductor \NCCO~(NCCO) has presented a number
of experimental challenges to the understanding of the physics
inherent to the system. Foremost among those is that, unlike other
cationically substituted systems, NCCO is not superconducting in
its as-grown state. Instead, a post-growth annealing procedure,
typically consisting of high-temperature exposure to an argon or
nitrogen atmosphere, is needed to induce superconductivity.
Ostensibly, this is necessary to remove excess oxygen impurities,
but the exact effect of the procedure is yet unclear.

Using X-ray and neutron diffraction as well as transmission
electron microscopy, we find that an unintended byproduct of this
reduction step is to partially decompose NCCO, forming layers of
oriented epitaxial neodymium cerium oxide, \NDO. Because of a
serendipitous match of lattice constants between NCCO and the
secondary phase, certain \NDO~ diffraction peaks are commensurate
with the NCCO reciprocal lattice. Upon application of a magnetic
field, \NDO~exhibits a paramagnetic response even at a temperature
of 1.9 K.  Knowledge about the structural and magnetic properties
of the secondary phase is essential in order to separate intrinsic
properties from extrinsic effects when investigating the
connection between magnetism and superconductivity in NCCO.
Specifically, the extensive data we present are inconsistent with
the notion of a field-induced quantum phase transition from a
superconducting to antiferromagnetic state of NCCO
\cite{NCCO:SecondPhase:DaiNature, NCCO:SecondPhase:DaiPRB,
NCCO:SecondPhase:Zhang} and have a more prosaic origin -- a
paramagnetic impurity phase.  A brief summary of some of our
results was published
recently.\cite{NCCO:SecondPhase:NatureComment}

This paper is organized as follows: after a discussion of the NCCO
crystal structure (Sec. II) and of experimental details (Sec.
III), we describe the oxygen-reduction procedure that is required
to render NCCO superconducting in the cerium-doping regime $0.13 <
x < 0.20$ (Sec. IV). We then summarize previous results for cubic
rare-earth (RE) oxides of the form RE$_2$O$_3$ (Sec. V) and
proceed to present our structural X-ray diffraction (Sec. VI) and
TEM (Sec. VII) work. Finally, we present detailed magnetic neutron
scattering results of the magnetic field effects observed in
reduced NCCO in Sec. VIII, and discuss our results together with
previous work in Sec. IX. A summary is given in Sec. X.

\section{Structure}

\NCCO~ crystallizes into a modified form of the body-centered
tetragonal K$_2$NiF$_4$ structure found in other single-layer
cuprate superconductors.\cite{NCCO:Structure:Radaelli} The
structure is tetragonal at all temperatures and is shown
schematically in Fig. 1.  What is unique about NCCO vis-$\rm
\grave{a}$-vis other single-layer cuprates is the nominal absence
of apical oxygen atoms.
\cite{NCCO:3phases:Goodenough,NCCO:3phases:Attfield,NCCO:3phases:Bringley}
Instead, the out-of-plane oxygen atoms are located directly above
and below the in-plane oxygens. This is known as the $T'$
structure.  In the more common $T$ structure, oxygen atoms are
present both above and below the copper site forming an octahedron
as in hole-doped \LSCO.  There exists a third, hybrid structure
($T^*$), in which one of the apical sites is occupied while the
other is vacant.

\begin{figure}[t]
\centering
\includegraphics{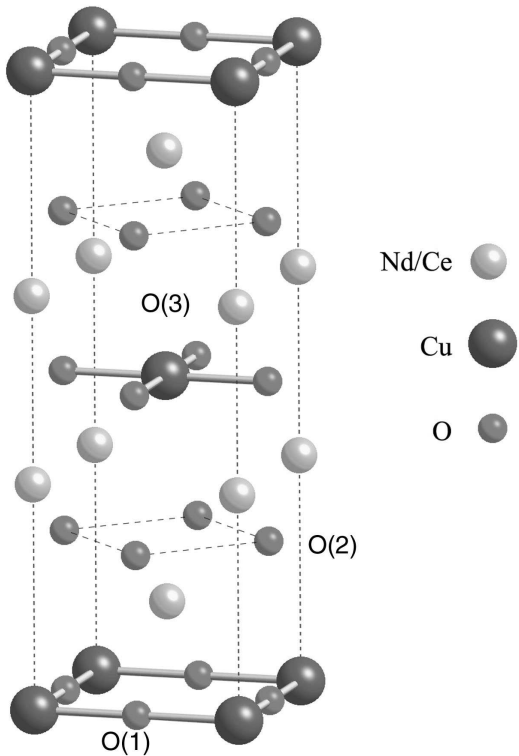} \caption{\label{NCCO
unit cell figure} Unit cell of \NCCO. The structure is of $I4/mmm$
(No. 139) symmetry and tetragonal at all temperatures. For the
undoped compound, $a=3.94$ \AA~ and $c=12.1$ \AA~ at room
temperature. The apical oxygen site O(3) is nominally vacant.  }
\end{figure}

A key characteristic of the $T'$ structure is the square-planar
coordination of the copper site. Apical oxygen disorder has been
put forward to explain why a reduction procedure is necessary to
produce a superconducting phase in NCCO. According to that
argument, the apical sites are randomly occupied by a small
fraction $\delta$ of interstitial oxygen in as-grown crystals. In
addition to changing the carrier density in the CuO$_2$ sheets,
this produces a random pinning potential for the doped electrons
which prohibits superconductivity. The purpose of the reduction
procedure is to remove the interstitial oxygen, after which
superconductivity may occur. Unfortunately, this simple picture
has been extremely difficult to verify. Generally,
non-stoichiometric oxide materials can be stable for a wide range
of oxygen concentrations
\cite{Book:NonStoichiometricOxides:Sorensen} and the precise
starting composition and oxygen site occupancies are typically
unknown.

Based on single-crystal neutron diffraction studies, Radaelli
\etal \cite{NCCO:Structure:Radaelli} concluded that the apical
O(3) site has an occupancy of $\sim 0.1$ at the position
$[0,0,0.2]$ in as-grown \NCO. In reduced samples, the occupancy
was found to be lowered to $\sim 0.04$. The authors also reported
less than full occupancy at both the O(1) in-plane and O(2)
out-of-plane sites, and these were not seen to vary beyond one
standard deviation with reduction.  Further neutron diffraction
experiments on NCCO ($x=0.15$) showed a smaller amount of apical
oxygen ($\sim 0.06$) in the as-grown state, which is lowered to
$\sim 0.04$ upon reduction.\cite{NCCO:Structure:Schultz} The
diffraction data indicate that interstitial oxygen is probably
present at the apical site in amounts negatively correlated with
the cerium doping level. Presumably, this anticorrelation is a
consequence of the contraction of the $c$-lattice constant,
\cite{LatticeConstants:Tarascon} which permits a decreasing number
of interstitials at higher doping.  The occupancies of the other
two oxygen sites were found to be similar to the undoped case
($x=0$). Because of the measured incomplete occupancy of the O(1)
and O(2) sites, even with the presence of additional interstitial
oxygen, summation over the oxygen site occupancies was found to
give stoichiometries of 3.97(4) and 3.95(3) for as-grown undoped
($x=0$) and doped ($x=0.15$) compounds, respectively. This
suggests that as-grown crystals may, in fact, be oxygen deficient.
In any case, the amount of oxygen removed, particularly in the
superconducting composition range ($0.13 < x < 0.20$), is near the
detection limit of neutron diffraction, further confounding
efforts to understand the exact effect of the reduction step.  A
M\"ossbauer study on $^{57}$Co substituted NCCO gave evidence for
the presence of apical oxygens even after reduction.\cite{nath94}
Other methods of measuring the oxygen content, such as
thermogravimetric analysis (TGA) and titration
techniques, typically are not site specific.

\section{Experimental}

All crystals used in this study were grown in the T.H. Geballe
Laboratory for Advanced Materials at Stanford University using the
travelling-solvent floating-zone (TSFZ) technique.  As-grown, NCCO
exhibits strong spin correlations in the paramagnetic phase and
long-range antiferromagnetic order at low temperature.  The
crystals were grown in 4 atm of O$_2$, which results in samples
with relatively high oxygen content and N\'eel temperatures.
\cite{mang03b} Cerium concentrations of the grown boules were
verified by Inductively Coupled Plasma (ICP) analysis. The cerium
solubility limit is close to $x=0.18$.  With the exception of the
$x=0.18$ sample, for which ICP carried out on different crystal
pieces indicated a somewhat inhomogeneous Ce distribution of $\pm
0.02$, the Ce concentrations were found to be uniform.  ``Optimal"
doping, with $T_{\rm c} = $ 24-25 K (onset), is achieved for
suitable reduction conditions and cerium concentrations of
$x=0.14-0.16$. Details pertaining to the samples used in the
present study are listed in Table I.

The room-temperature in-plane lattice constant of NCCO, $a \approx
3.94$ \AA, increases very slightly with doping, by about 0.3\%.
The $c$-lattice constant decreases by about 1\% with cerium
doping, from 12.16 \AA~in the undoped material to 12.05 \AA~for
$x=0.18$. In this paper, we will index reflections according to
two different schemes. One is based on the tetragonal unit cell of
NCCO. The other is based on the cubic unit cell of
Nd$_2$O$_3$.\cite{Er2O3:Moon} Reflections indexed according to the
latter scheme will be marked by the subscript ``c."

X-ray diffraction data were taken in reflection mode at beam line
7-2 of the Stanford Synchrotron Radiation Laboratory. The beam was
monochromatised at 14 keV from a wiggler spectrum using a Si(111)
double-crystal monochromator. The X-ray attenuation length was
approximately 20 $\mu$m. Neutron diffraction data were taken using
the spectrometers BT2 at the National Institute of Standards and
Technology, E4 at the Hahn-Meitner-Institut in Berlin, Germany,
and N5 at Chalk River Laboratories in Chalk River, Canada.
Transmission electron microscope (TEM) images were taken with a
FEI CM20 FEG-TEM instrument in the T.H. Geballe Laboratory for
Advanced Materials at Stanford University.

\begin{figure}[b]
\centering
\includegraphics{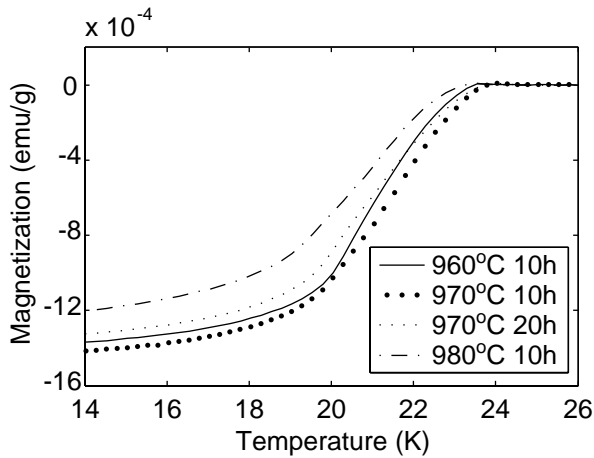}
\caption{\label{AnnealFigure} AC magnetic susceptibility of
Nd$_{1.86}$Ce$_{0.14}$CuO$_{4 \pm \delta}$ crystals reduced in
argon under various conditions. A subsequent low-temperature
oxygen anneal (not shown) raises the transition temperature by
approximately 1 K. The stated temperature is the temperature at
the sample position. The control temperature of the tube furnace
is measured by an external thermocouple, situated outside the
quartz tube at the bottom of the sample chamber. There is a
difference of 38 K between the control temperature and the
(higher) sample temperature which is measured by a thermocouple
placed directly above the sample inside the quartz tube.
\cite{TemperatureNote}  Susceptibility measurements were
performed using a Quantum Design PPMS system in an AC excitation
field of 0.02 Oe at 10,000 Hz.}
\end{figure}

\section{Reduction}

Regardless of the microscopic effects of the reduction step, it is
clear that such a procedure is necessary to induce
superconductivity. Kim and Gaskell
\cite{PhaseStability:KimGaskell} investigated the stability field
of Nd$_{1.85}$Ce$_{0.15}$CuO$_{4 \pm \delta}$ as a function of
temperature and oxygen partial pressure. They determined that
\NCCO~ is unstable towards decomposition into \NDO~ +
NdCeO$_{3.5}$ + Cu$_2$O under sufficiently severe reducing
conditions. Interestingly, they found that superconductivity only
appears for conditions in which CuO is converted to Cu$_2$O in the
binary system Cu-O, and that the superconducting transition
temperature monotonically increases as one approaches the phase
decomposition
boundary during the reduction process.

\begin{table*}
\caption{\label{SampleTable} Table of the reduced samples studied,
with cerium concentration, reduction conditions, experimental
probe used, approximate \NDO~volume fraction (VF),\cite{VF} as
well as estimates of the average thickness of the secondary phase
along [0,0,1].}

\begin{minipage}{\textwidth}
\begin{ruledtabular}
\begin{tabular}{cccccc}
  Sample & x & Reduction Condition & Probe & \NDO~ VF & Thickness (\AA) \\
  \hline
  NCCO0\footnotemark[1] & 0 & 890\degC/11 h/Ar & X-rays & 0.02\%& - \\
  NCCO10\footnotemark[2]& 0.10 & 960\degC/20 h/Ar; 500\degC/20 h/O$_2$ & neutrons & 2\% & 100 \\
  NCCO14LR & 0.14 & 960\degC/1 h/Ar & X-rays & 0.02\% & 50 \\
  NCCO14& 0.14 & 960\degC/20 h/Ar; 500\degC/20 h/O$_2$ & X-rays & 1\%&  70\\
  Core/Shell\footnotemark[3] & 0.14 & 970\degC/10 h/Ar; 500\degC/20 h/O$_2$ & neutrons, X-rays & 0.5\% & 70 \\
  NCCO16\footnotemark[4] & 0.16 & 960\degC/20 h/Ar; 500\degC/20 h/O$_2$ & TEM & 2\% & 60 \\
  NCCO18 & 0.18 & 960\degC/10 h/Ar & neutrons & 0.5\% & 140 \\
\footnotetext[1]{Insufficient data to determine thickness.}
\footnotetext[2]{Thickness determined from magnetic measurements
of the \NDO~secondary phase (from fits shown in Fig. 13).}
\footnotetext[3]{Volume fraction and thickness are from X-ray
diffraction on a small piece of crystal, prior to separating the
sample into core/shell pieces.}
\end{tabular}
\end{ruledtabular}
\end{minipage}
\end{table*}

\begin{figure}[b]
\centering
\includegraphics{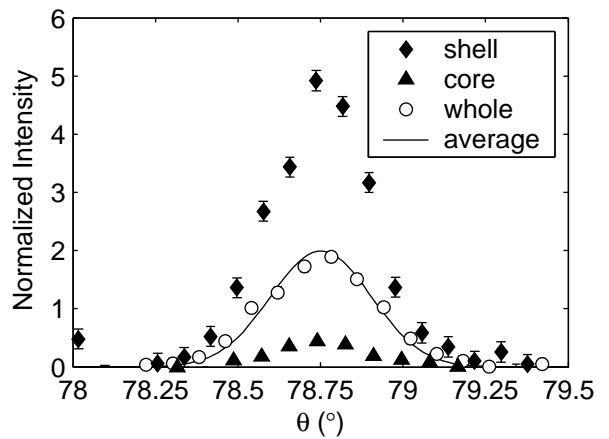}
\caption{\label{CorevShell} Neutron diffraction mosaic scans
through the antiferromagnetic (1/2,1/2,1) peak for an
oxygen-reduced $x=0.14$ sample (Core/Shell; see Table 1).
Individual scans are shown for the outer shell, the inner core,
and the whole crystal prior to division. The solid line represents
a weighted average of the shell and core signals. Intensities were
normalized by counting time, sample mass, and neutron flux. The
data were taken at the NIST Center for Neutron Research.}
\end{figure}

Our own studies indicate that the resultant superconducting state
is highly dependent on the parameters used in the reduction
process.  Figure \ref{AnnealFigure} demonstrates the sensitivity
of the transition temperature and the magnetic susceptibility on
the annealing conditions. In order to find optimal conditions,
that is, conditions that result in a sample with the highest \Tc,
we employed a tube furnace in which the sample was completely
enclosed in a quartz tube and could be rapidly heated and quenched
in a controlled atmosphere. We used high-purity argon gas
containing less than 0.5 ppm O$_2$. A series of small samples of
approximately 400 mg were cut from a single-crystal boule of NCCO
($x=0.14$) and were exposed to an argon environment at different
temperatures and for varying durations.\cite{TemperatureNote} A
measurable change in sample mass, ranging from 0.1\% to 0.3\% was
observed, corresponding to a change in oxygen content on the order
of $\delta=0.04$. Samples annealed at 990\degC~ and above suffered
severe damage and eventually decomposed. Examination of the
reduced samples revealed that the interior of the sample was less
damaged than the exterior. We find the optimal annealing
conditions for $x=0.14$ to be 970\degC~ for 10 h in argon,
followed by 500\degC~ for 20 h in oxygen. In general, the
additional low-temperature oxygen anneal tends to increase \Tc~by
approximately 1 K. Other work \cite{NCCO:Superstructure:Kurahashi}
indicates the existence of two metastable phases for $x=0.15$,
with values of \Tc~ of 18 K and 25 K. The higher-\Tc~samples were
produced under conditions similar to ours, whereas the
lower-\Tc~ones were created in a more benign reducing environment
with an oxygen concentration of 1000 ppm.

\begin{figure}[t]
\centering
\includegraphics[height=10cm]{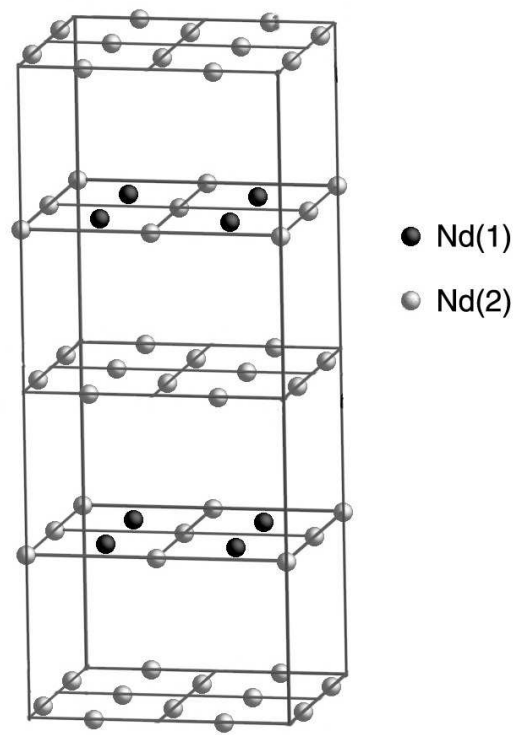}
\caption{\label{CubicNd2O3} Stylized Nd$_2$O$_3$ unit cell,
modeled after
  Fig. 2 of Ref. \onlinecite{Er2O3:Moon}. The unit cell is cubic, but the
  $c$-axis has been expanded and the oxygen atoms are not shown for
  clarity. There are two crystallographically distinct Nd sites; the ones
  in grey have C$_2$ site symmetry and the ones in black have C$_{3i}$ symmetry.
  The C$_2$ atoms are displaced from the shown positions
  along one of the cell edges
  by about 1/30 of a unit cell.
  }

\end{figure}

The difference we observed between the interior and exterior
regions of our samples suggests that, because of the long time
scale for oxygen diffusion, the resultant oxygen distribution in
large reduced samples is not homogeneous. In order to test this
proposition, we reduced a cylindrical crystal boule ($x=0.14$) and
cut it into two halves, perpendicular to the cylinder axis. One
half was mechanically ground down to remove the surface. This
decreased the diameter from 4.2 mm to 3.2 mm and revealed the
inner core of the crystal. The other half was cut lengthwise and
the inner section was removed to create a hollow shell that was 1
mm thick. Transverse scans through the (1/2,1/2,1) magnetic Bragg
reflection were taken for both the core and the shell pieces and
are compared in Fig. \ref{CorevShell} to a scan previously made on
the entire boule. The magnetic signal from the shell piece is an
order of magnitude stronger than that of the core, but the
averaged signal of the two contributions, weighted by the mass of
each section, essentially reproduces the signal of the whole
sample. Due to the weakness of the signal, it is difficult to
assess the exact onset of magnetic order in the core piece, but
the \Neel~ temperature ($T_{\rm N} \sim 100$ K) does not likely
differ by more than 20 K between the two pieces. This suggests
that the core is in some sense the better sample, since it has an
equally high superconducting transition temperature and a weaker
antiferromagnetic phase. Remarkably, despite the large difference
in the strength of the magnetic signal, no difference can be
detected in the onset of superconductivity between the core and
the shell piece. It is difficult to obtain accurate estimates of
superconducting volume fractions from magnetic susceptibility
data, and such measurements, performed on pulverized pieces of the
core and shell sample, gave superconducting volume fractions of
$10(5)\%$. It is possible that the reducing conditions employed
are too severe, and that chemical decomposition destroys
superconductivity beyond a certain threshold. The core piece,
protected by the shell, is less exposed to the severe reducing
atmosphere. This is consistent with work by Brinkmann \etal
\cite{NCCO:Reduction:Brinkmann} who reported an extended
superconducting dome down to $x=0.08$ in NCCO when samples were
subjected to a reduction temperature of 1000\degC, but protected
from decomposition by being sandwiched between chemically
homogeneous polycrystalline slabs.

\begin{figure}[t]
\centering
\includegraphics{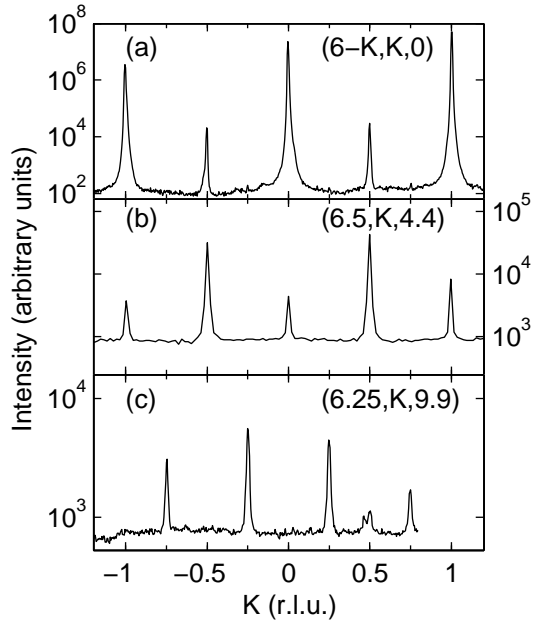}
\caption{\label{SSRL1} X-ray diffraction data for a reduced
$x=0.14$
  sample (NCCO14; see Table I). The scans were taken at (a) 30 K, (b) 300 K,
  and (c) 7 K. Note the logarithmic intensity scale.
  Magnetic susceptibility measurements give \Tc = 24 K (onset).}
\end{figure}

\begin{figure}[t]
\centering
\includegraphics{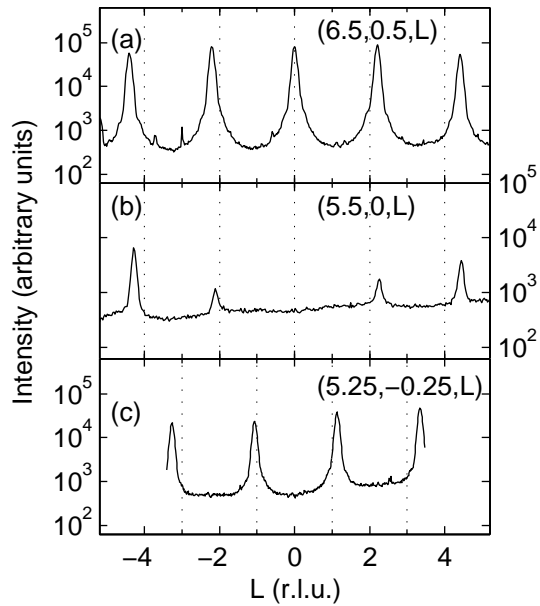}
\caption{\label{SSRL2} Scans perpendicular to the copper-oxygen
plane for
  the same sample as in Fig. \ref{SSRL1} (NCCO14; see Table 1), taken at
  room temperature. Note the logarithmic intensity scale.}
\end{figure}

\section{\label{RareEarthOxides}Rare Earth Oxides}

Rare-earth elements form polymorphic sesquioxides of the form
RE$_2$O$_3$. There exist three low-temperature (A,B, and C) and
two high-temperature (H and X) structural
phases.\cite{RareEarthReview} The preferred phase varies with
rare-earth ionic radius and temperature. In the case of
Nd$_2$O$_3$, the trigonal A-modification is most commonly found. A
second, cubic variant, the C-modification, has also been observed.
The cubic phase has the bixbyite structure with space group $Ia3$
(No. 206).\cite{Book:Wyckoff,Book:CrystallographyTables} The unit
cell is very large, with 32 rare-earth ions occupying two
different crystallographic sites. Because of the difficulty in
preparing the cubic modification of Nd$_2$O$_3$ its magnetic
structure is unknown, to the best of our knowledge. However, Moon
\etal\cite{Er2O3:Moon} studied the magnetic structures of
Er$_2$O$_3$ and Yb$_2$O$_3$, which also possess the bixbyite
structure.  The two systems were found to have N\'eel temperatures
of $T_{\rm N} = 3.4$ K and 2.3 K, respectively, and differing
antiferromagnetic structures, each describable by four magnetic
sublattices. Complex magnetic field effects were observed for
Er$_2$O$_3$, but not reported in detail.

\section{\label{X-ray}Structural X-ray Diffraction}

Additional diffraction peaks in reduced samples of NCCO have been
observed previously by both
electron\cite{NCCO:Superstructure:Izumi} and
neutron\cite{NCCO:Superstructure:Kurahashi} diffraction. These
additional peaks have been attributed, respectively, to the
formation of an oxygen vacancy superstructure and to coherent
atomic displacements. Using X-ray diffraction, we investigated the
formation of these additional peaks at two cerium concentrations:
undoped ($x=0$) and optimally-doped ($x=0.14$). Additional data
for a lightly-reduced $x=0.10$ sample (not shown) are consistent
with the results presented here.

The additional diffraction peaks are not observed in as-grown NCCO
at any cerium concentration, but are present in all samples that
have been subjected to a reduction treatment, including the
undoped sample. The peaks appear relatively early in the reduction
process, and were observable in a sample that had been reduced for
only one hour, compared to the 10-20 h normally employed to obtain
optimized superconducting properties.  The samples exhibited
varying degrees of diffuse scattering, examples of which will be
given shortly. The sample with the sharpest additional peaks and
the least diffuse scattering among all crystals studied with
X-rays was an optimally-doped NCCO ($x=0.14$) crystal, which we
discuss first.

From Figs. \ref{SSRL1} and \ref{SSRL2} we identify three classes
of additional peaks. Figure \ref{SSRL1}a reveals both the
anticipated Bragg peaks from the NCCO structure at integer $K$
positions and an additional contribution at half-integer
positions, such as (5.5,0.5,0). The latter are located at the same
position in reciprocal space as the antiferromagnetic zone center,
but are unconnected with magnetism in NCCO. Non-resonant X-ray
scattering is only weakly sensitive to magnetism, and such peaks
are visible at room temperature, well above the reported magnetic
ordering temperature. The scan shown in the middle panel reveals a
second set of half-integer peaks, such as (6.5,0,4.4), that occur
when one in-plane index is an integer. The last panel reveals a
final set of peaks occurring at quarter-integer positions, such as
(6.25,0.25,9.9).

Although these peaks appear at rational $H$ and $K$ indices, they
possess an unusual $L$-dependence. Careful examination of Fig.
\ref{SSRL2} reveals that, for these three classes of peaks, the
incommensurability systematically increases along the out-of-plane
direction. This is different from scattering due to a
superstructure, in which case the incommensurability is constant
for all Brillouin zones. The additional diffraction peaks have $L$
indices of \{0,$\pm$1.1,$\pm$2.2,\ldots\}. This peculiar
$L$-dependence suggests that these peaks are due to a secondary
phase that is well oriented with the [001] surface of NCCO, and
that the $c$-lattice constant is approximately 10\% smaller than
that of NCCO.

The widths of the additional diffraction peaks are anisotropic
with respect to the in-plane and out-of-plane directions. For
example, while the half-width at half-maximum (HWHM) of the
(5.5,0.5,2.2) peak along the in-plane scan direction ($\sim0.0043$
r.l.u.) is only slightly larger than that of the NCCO Bragg peaks,
the width along the out-of-plane direction ($\sim0.066$ r.l.u.) is
an order of magnitude larger than for the NCCO Bragg peaks. As
discussed in more detail below, this implies an in-plane
correlation length \xiab~of at least several hundred \AA, an
out-of-plane correlation length \xic~of only 30 \AA, and a layer
thickness of approximately 70 \AA.

\begin{figure}[tb]
\centering
\includegraphics{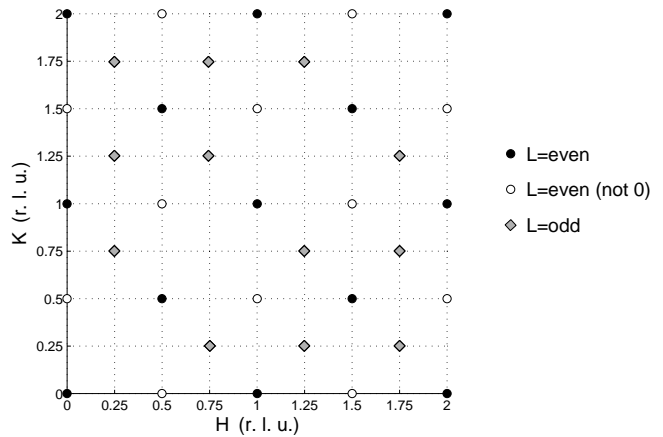}
\caption{\label{map} ($H,K,L$) zone map of reciprocal space
indicating the   cubic \NDO~positions where peaks are observed in
reduced samples of   NCCO.  $L = \text{even}$ indicates that
out-of-plane, peaks are centered   at even integer multiples of $L
= \pm 1.1$: $L=\{0,\pm2.2,\pm4.4,\ldots\}$. $L = \text{even}$ (not
zero) implies the   same, except that no peak is observed for
$L=0$. $L = \text{odd}$   indicates that peaks are centered at $L
= \{\pm1.1,\pm3.3,\ldots\}$.   Note that this map does not include
the reciprocal space positions where scattering becomes allowed
once the glide-plane symmetry of cubic \NDO~is broken (see also
Fig. 8). }
\end{figure}

The width of the diffraction peaks emanating from the secondary
phase can be used to estimate the average thickness of the
epitaxial layers of \NDO. A lower-bound estimate is provided by
converting the Gaussian half-width at half-maximum (HWHM) into an
approximate correlation length using the critical scattering
formula $\xi=$1/HWHM(\AA$^{-1}$). The thickness might then be
considered to be approximately twice the correlation length. An
upper-bound estimate is provided by using the formula for
diffraction from an $N$-slit grating: $\sin(N\pi z)/\sin(\pi z)$,
where $N$ is the number of unit-cell layers of the secondary phase
and $z$ is expressed in reciprocal lattice units.
\cite{SecondaryPhase:N-SlitFormula} The two functional forms
provide estimates of the thickness that differ by approximately
40\%. The $N$-slit formula assumes that the epitaxial layers are
of uniform thickness, and also implies the existence of secondary
diffraction maxima (see, for example, Fig. 5 of Ref.
\onlinecite{SecondaryPhase:N-SlitFormula}), which we do not
observe for \NDO. Numerical simulations done assuming that the
widths of the epitaxial layers are sampled from a distribution,
and then averaging over the resultant signal to smooth out the
secondary maxima, indicate that the $N$-slit formula overestimates
the average layer thickness by approximately 20\%. On the other
hand, the Gaussian formula underestimates it by approximately
20\%. In this manuscript, the correlation length $\xi_{\rm c}$
along [0,0,1] is estimated using the Gaussian formulism. However,
the layer thickness obtained from diffraction, as reported in
Table I, is corrected by a factor of 2.4 to account for both the
20\% underestimation and the conversion from correlation length
(half-thickness) to thickness.




On the basis of this extensive survey, we construct the map of
reciprocal space in Fig. \ref{map}. A new unit cell can be drawn
with dimensions ($2\sqrt{2}a,2\sqrt{2}a,c/1.1$), where $a$ and $c$
refer to the lattice constants of the tetragonal NCCO unit cell.
We find that this is consistent with the presence of an additional
phase of cubic \NDO~ that is epitaxially oriented along the
diagonal direction of the CuO$_2$ plane. Because they bear no
special relationship with the surrounding matrix, secondary phases
typically form polycrystalline powder inclusions that result in
rings of scattering intensity at constant momentum transfer.
However, in the present case, the nearly perfect ($\sim 0.5\%$)
match between the lattice constant of \NDO~($a_{\rm c} \sim11.08$
\AA) and $2\sqrt{2}a$ ($\sim11.14$ \AA) leads to the formation of
a quasi-two-dimensional structure that is well oriented with the
copper-oxygen plane, but extends only a few unit cells along
[0,0,1]. We note that the cubic lattice constant $a_{\rm c}$ is
$\sim 10\%$ smaller than the $c$-axis
lattice constant of NCCO.

\begin{figure}[t]
\centering
\includegraphics{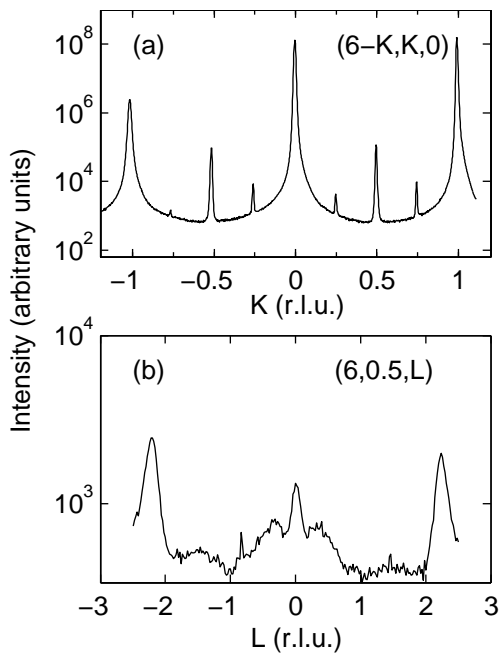}
\caption{\label{SSRL3} Room-temperature data for a lightly-reduced
$x=0.14$
  sample of NCCO (NCCO14LR; see Table I). Note the logarithmic intensity scale.
 (a) A cut across diffuse rods centered above and below the plane of the scan at $L=\pm
  1.1$ results in the appearance of pseudo-peaks at quarter positions that
  were not visible in Figs. \ref{SSRL1} and \ref{SSRL2}. (b) Large
  amounts of diffuse scattering not seen in Fig. \ref{SSRL2} culminating
  in a peak at $L=0$. This is evidence that the glide-plane symmetry of
  \NDO~ is broken.}
\end{figure}

\begin{figure}[t]
\centering
\includegraphics{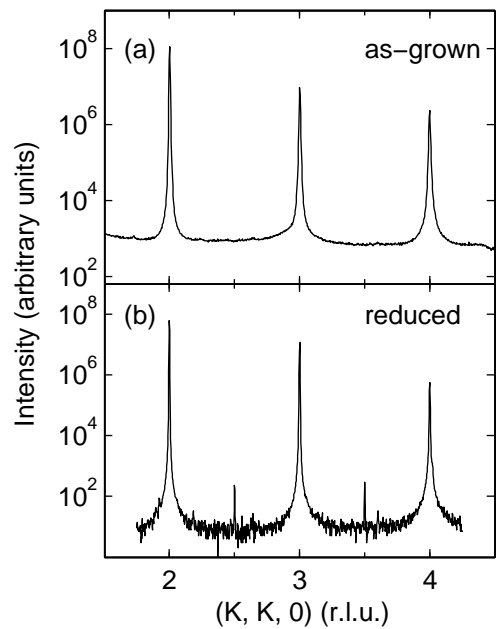}
\caption{\label{SSRLPure} Room-temperature scan along $(K,K,0)$ in
an
  undoped sample ($x=0$) of NCCO (NCCO0; see Table I). Note the logarithmic
  intensity scale. (a) The sample is in its as-grown state and no
  scattering is visible at the half-integer positions. (b) After
  being reduced in flowing Ar at 890\degC~for 11 h, new peaks appear at
  the half-integer positions.  }
\end{figure}




The secondary phase is strained by the surrounding NCCO matrix,
which causes the lattice constants to be significantly different
between the $c$-direction (10.95(2) \AA) and the other two axes
(11.14(2) \AA). However, the unit cell volume of the strained
compound is equal to that of bulk Nd$_2$O$_3$ ($a_{\rm c} = 11.08$
\AA):\cite{RareEarthReview} $11.14^2~ \text{\AA}^2 \times 10.95~
\text{\AA} \approx 11.08^3~ \text{\AA}^3$.  Hence, although the
structure of the quasi-two-dimensional epitaxial decomposition
phase is related to that of bulk Nd$_2$O$_3$, the symmetry is
reduced from cubic (space group $Ia3$) to orthorhombic.
Furthermore, there is evidence that the glide-plane symmetry
present in the original space group ($Ia3$) is broken. In samples
with relatively low volume fractions of the secondary phase, weak
additional scattering was easily observed at positions forbidden
by the glide-plane symmetry. In those cases, the diffuse
scattering is intense enough to result in a small peak at
(6,0.5,0) (equivalent to (13,11,0)$_{\rm c}$), as shown in the
lower panel of Fig. \ref{SSRL3}, a position forbidden by the
glide-plane symmetry. We note that this central peak has the same
$L$-width as the two peaks at $(6,0.5,\pm2.2)$. The large amount
of diffuse scattering is also responsible for what appear to be
peaks at quarter-integer positions in the upper panel of Fig.
\ref{SSRL3}. This intensity actually results from bisecting a rod
of diffuse scattering that connects the quarter-integer peaks
located above and below the scattering plane at $L=\pm1.1$. The
relative amount of diffuse scattering present in the two $x=0.14$
samples can be gauged by comparing Figs. \ref{SSRL1}a and
\ref{SSRL3}a,
and by comparing Fig. \ref{SSRL2}b with Fig. \ref{SSRL3}b.

In Fig. \ref{SSRLPure}, we compare a scan along $(K,K,0)$ for an
undoped sample ($x=0$) in its as-grown state, where no scattering
is visible at the half-integer positions, with data taken after
reduction, which reveal new peaks at the half-integer positions.
This sample was reduced at a lower temperature (890\degC~compared
to 960-970\degC~for the Ce-doped samples), because undoped samples
completely decompose after being reduced at 960\degC. Note that
the peaks here are approximately two orders of magnitude smaller
than in the previous cases, resulting in a \NDO~ volume fraction
estimate of only 0.02\%.




\section{\label{TEM}Transmission Electron Microscopy}

\begin{figure}[tbh]
\centering
\includegraphics{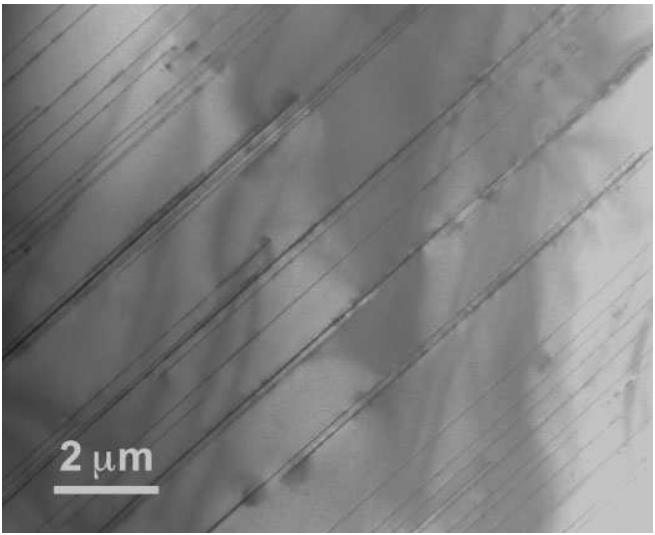}
\caption{\label{TEM1} Medium-resolution [1,0,0] TEM image of a
reduced $x=0.16$ sample (NCCO16; see Table 1) with layers of the
\NDO~phase perpendicular to [0,0,1]. }
\end{figure}

To confirm our diffraction results with real-space information, we
used transmission electron microscopy (TEM) to characterize the
microstructure of a reduced $x=0.16$ crystal.  A small section of
the crystal was prepared with a [1,0,0] surface, and the $b$-$c$
plane was imaged.  Figure \ref{TEM1} shows an image that reveals
thin straight layers of the secondary phase perpendicular to
[0,0,1] of NCCO. The secondary-phase regions have a spatial extent
well above 1 $\mu m$ parallel to the CuO$_2$ planes of
NCCO, i.e., perpendicular to [0,0,1].

\begin{figure}[tbh]
\centering
\includegraphics{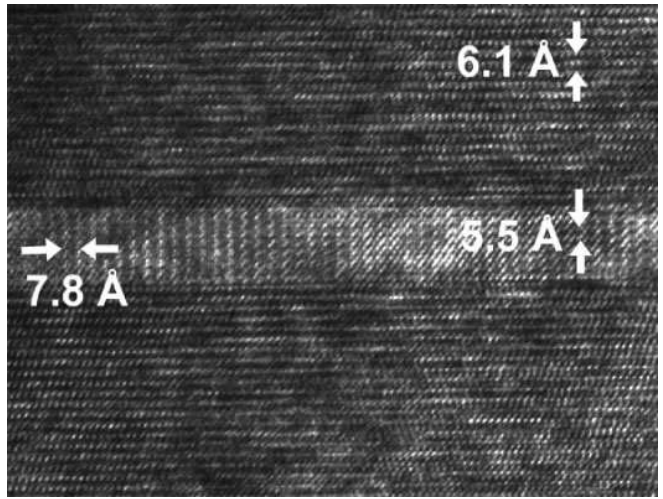}
\caption{\label{TEM2} High-resolution TEM image of the same
crystal as in
  the previous figure with a secondary phase region of typical width,
  showing the epitaxial orientation of \NDO~with the NCCO matrix.  The 6.1
  \AA~(002) planes of the matrix are parallel to the 5.5 \AA~(002)$_{\rm
    c}$ planes of the secondary phase.  The 7.8 \AA~perpendicular spacing
  for the secondary phase represents the (110)$_{\rm c}$ planes, which are
  parallel to the (020) planes of the matrix.  }
\end{figure}

In Fig. \ref{TEM2}, we show a high-resolution image of a
secondary-phase layer of typical width.  \NDO~has an epitaxial
relationship with the tetragonal NCCO matrix of $[0,0,1]_{\rm c}
\parallel [0,0,1]$ and $[1,1,0]_{\rm c} \parallel [100]$. The
width of the \NDO~layers is minimized in the direction of greatest
mismatch, i.e., along the $c$-axis of the NCCO.  While there are
some rare regions as thick as 150 \AA~and some regions thinner
than 40 \AA, by averaging over 40 \NDO~layers identified on a
series of photographs which spanned different regions of the
sample, we estimate that the typical thickness is in the range
40-80 \AA. This is in very good agreement with our diffraction
results listed in Table I.  Again using information from several
images such as that of Fig. \ref{TEM1}, we arrive at a crude
estimate of a secondary phase volume fraction of $1-3\%$. We
conclude that our complementary real-space TEM results are
entirely consistent with the momentum-space
information presented in the previous Section.

\section{\label{MagFieldEffect}Magnetic Field Effects}

Having established the presence of \NDO~as a result of the
decomposition of the NCCO matrix during reduction, we now
investigate how this secondary phase responds to the application
of a magnetic field. This is important because peaks from the
secondary phase are found at positions associated with the
antiferromagnetic zone center of NCCO. Therefore, it is essential
to know how \NDO~responds in order to avoid confusion between
intrinsic and extrinsic effects when investigating the connection
between magnetism and
superconductivity.

For this purpose, we placed a reduced, superconducting sample of
NCCO ($x=0.18$, \Tc~$\sim 20$ K) in the horizontal-field magnet M2
at Chalk River Laboratories.  The sample was 1.6 cm long, 3.6 mm
in diameter, and weighed 1.28 g.  Because of a large easy-plane
anisotropy for magnetic neodymium atoms in NCCO, the application
of a magnetic field along [0,0,1] exerts a torque on the sample
that attempts to twist it until the copper-oxygen planes are
oriented parallel to the field direction. To counteract this,
opposite faces of the sample perpendicular to the $c$-axis were
ground flat and placed into a press-shaped sample mount.  The
sample was held fast in the mount by tightening the mount with
screws against the flat sections of the sample.

\begin{figure}[h]
\centering
\includegraphics{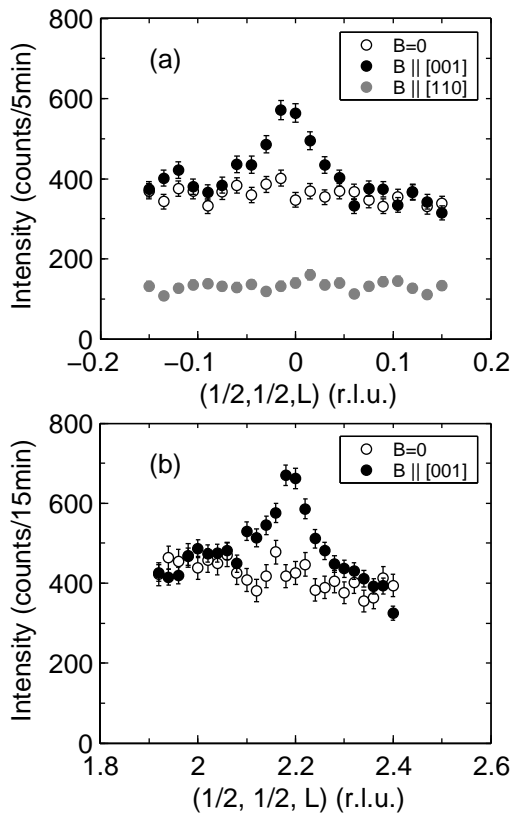}
\caption{\label{ChalkPiPi} $L$-scans through (a)
  (1/2,1/2,0) and (b) (1/2,1/2,2.2). The magnetic
  field strength is either 0 or 2.6 T.  The sample is a $x=0.18$ crystal
  with \Tc = 20 K (NCCO18; see Table 1). The data were taken at Chalk
  River Laboratories at a temperature of 2 K.}
\end{figure}

At zero field, strong peaks from the secondary phase are visible.
By comparing the integrated intensities of the (4,0,4)$_{\rm c}$
reflection of cubic \NDO~(the (1,1,4.4) reflection in the NCCO
reciprocal lattice) to the (0,0,2) and (0,0,4) reflections of
NCCO, we estimate the volume fraction of the secondary phase to be
approximately 0.5\% in this sample.

In Fig. \ref{ChalkPiPi}a, we present $L$-scans through
(1/2,1/2,0), also indexable as (2,0,0)$_{\rm c}$. This position is
associated with the antiferromagnetic wavevector of the CuO$_2$
plane. Weak reflections are not always easily visible in neutron
measurements due to the relatively low flux of neutron sources
when compared to X-ray synchrotrons as well as the relatively high
background scattering and coarse momentum resolution in neutron
scattering experiments.  In the present case, no peak can be
discerned above background at zero field, as expected, given the
relatively small sample size and secondary phase volume fraction
when compared to our $x=0.10$ sample for which we do observe a
weak structural peak (Fig. 13a). With the application of a
magnetic field of 2.6 T directed along [0,0,1], perpendicular to
the copper oxide planes, a peak is readily visible.  In Fig.
\ref{ChalkPiPi}b, we present a similar set of scans through the
position (1/2,1/2,2.2), or (2,0,2)$_{\rm c}$. As in the case of
the (2,0,0)$_{\rm c}$ reflection, no peak is visible at zero
field, but upon application of a 2.6 T magnetic field directed
along [0,0,1], a peak clearly emerges at $L=2.2$.

From the extensive X-ray survey of Sec. VI, we know that these two
positions are associated with structural peaks of a secondary
decomposition phase of \NDO. The application of a magnetic field
polarizes the Nd moments of the secondary phase, which results in
a ferromagnetic component that enhances the scattering at nuclear
positions of the secondary phase, and renders the (2,0,0)$_{\rm
c}$ and (2,0,2)$_{\rm c}$ peaks visible. This point is well
demonstrated by Fig. \ref{ChalkPiPi}b, as the position
(1/2,1/2,2.2) is manifestly unrelated to the NCCO structure.  We
test the proposition that the field-induced magnetism is
ferromagnetic in nature by studying the effect of a 2.6 T field
along [1,1,0] for scattering at the (1/2,1/2,0) position.  As can
be seen from Fig. \ref{ChalkPiPi}a, we find that a peak is no
longer visible. This is due the geometric factor $S_\perp \equiv
\hat{\textbf{Q}}\times({\textbf{S}}\times\hat{\textbf{Q}})$ in the
magnetic neutron scattering cross-section for unpolarized
neutrons,\cite{Book:Scattering:Shirane}
\begin{equation}\label{MagCrossSection}
 \frac{d\sigma}{d\Omega_f}=N\frac{(2\pi)^3}{V}\left|\sum_j\left(\frac{g\gamma r_0}{2}\right)
S_\perp e^{i\textbf{Q}\cdot\textbf{r}_j}\right|^2,
\end{equation}
which implies that the observed intensity is proportional to the
square of the component of the moment perpendicular to
$\textbf{Q}$.  Here, $N$ is the number of unit cells, $V$ is the
volume of one unit cell, $g$ is the Land\'{e} factor, $\gamma$ is
the gyromagnetic ratio, $r_0$ is the classical electron radius,
and $\textbf{r}_j$ are the atomic positions in the unit cell.
Because the Nd moments are field-induced, their direction tends to
follow the direction of the applied field. Neglecting possible
small anisotropy effects, when the magnetic field is oriented
along [0,0,1], the moment is perpendicular to the scattering
wavevector $\textbf{Q}=(1/2,1/2,0)$ and the maximum contribution
is obtained. However, when the field is directed along [1,1,0],
the moments are parallel to the field, and hence parallel to
$\textbf{Q}=(1/2,1/2,0)$. In this geometry, the magnetic
contribution to the scattering is zero and no peak can be
distinguished. The increased background level in some scan
geometries occurs for a narrow region of the magnet where the
electrical wiring and cryogenic tubing is located, which leads to
higher incoherent scattering than neighboring parts of the magnet
when illuminated by the incident neutron beam.

\begin{figure}[t]
\centering
\includegraphics{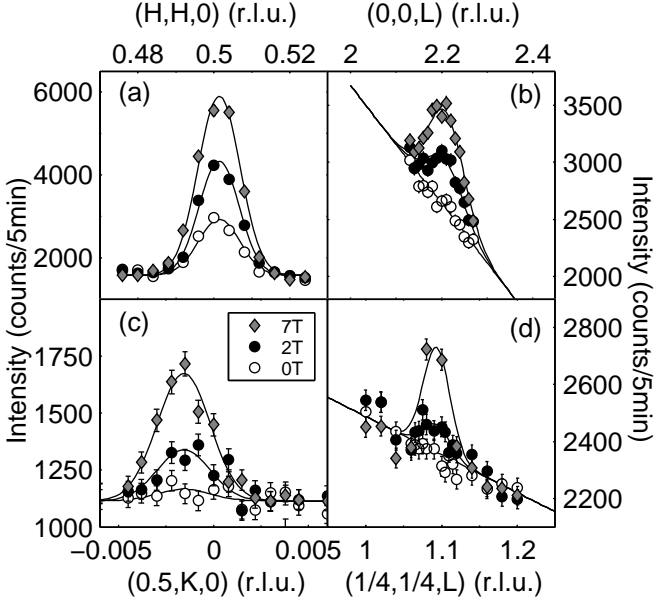}
\caption{\label{10NSC} Magnetic neutron diffraction scans for a
reduced
  $x=0.10$ \textit{non-superconducting} sample (NCCO10; see Table 1).
  In all cases, care was taken to ensure that the magnetic field was applied
  along a cubic \NDO~ axis and perpendicular to the scattering wavevector.
  Only scan (b) was originally taken for 5 min/point.
  All other scans were normalized to
  5 min/point for clarity.
  (a,b) Scans through (1/2,1/2,0)
  ((2,0,0)$_{\rm c}$) and the cubic-equivalent position (0,0,2.2)
  ((0,0,2)$_{\rm c}$). (c,d) Scans through (1/2,0,0) ((1,1,0)$_{\rm
  c}$) and the cubic-equivalent position (1/4,1/4,1.1) ((1,0,1)$_{\rm
  c}$).  The large sloping background for the (0,0,2.2) and (1/4,1/4,1.1)
  reflections results, respectively, from the
  close proximity to the
  strong NCCO (0,0,2) Bragg reflection and the proximity to the
  direct beam at (0,0,0). This prohibits the observation of
  zero-field \NDO~structural scattering. However, from fits of the
  field-induced data (solid lines) we extract a magnetic correlation
  length of $\xi_{\rm c} \approx 40$ \AA, consistent with the structural
  results reported in Table I.  The data were taken at the NIST Center for Neutron
  Research.  }
\end{figure}

Since the origin of the field-induced scattering is the secondary
decomposition phase that forms as a result of the oxygen-reduction
procedure, it is naturally not observable in as-grown samples.  Of
course, this does not imply that the observed field effect is
related to superconductivity in the NCCO matrix (which occurs for
$x>0.13$), but only that reduction is a necessary condition to
form the secondary phase as well as to induce superconductivity.
To demonstrate this, Fig. \ref{10NSC} shows magnetic scattering in
an oxygen-reduced, {\em non-superconducting} NCCO crystal
($x=0.10$). The data were taken at the NIST Center for Neutron
Research. The sample measured 4 mm in diameter, 2.1 cm in length,
and weighed 3.5 g. The sample volume, the volume fraction of the
secondary phase ($\sim 2 \%$), and the neutron flux are larger
than in the case of the $x=0.18$ sample. As a result, a structural
peak is visible at (1/2,1/2,0) (i.e., (2,0,0)$_{\rm c}$) in zero
field. We also demonstrate enhanced scattering in a magnetic field
at the equivalent position (0,0,2.2) (i.e., (0,0,2)$_{\rm c}$).
We conclude that an enhancement of scattering in a magnetic field
is observable in reduced NCCO, \textit{irrespective} of whether
the sample is superconducting or not.  Furthermore, the effect is
observable at any position corresponding to a Bragg-like
reflection of the secondary \NDO~ phase, subject, of course, to
the modulation of the structure factor and the geometric factor
$S_\perp^2$ for that particular reflection.

An important point is that although the unit cell of the secondary
phase is cubic, it is not simple cubic. As discussed in Sec.
\ref{RareEarthOxides}, cubic RE$_2$O$_3$
has two crystallographically distinct
rare-earth sites. In principle, each site may have a distinct
magnetic moment. As the separate contributions to the structure
factor from each site add destructively for some reflections, the
observed intensity for a given reflection is a function of the
relative strength of the magnetic moments on the two sites. For
cubic Nd$_2$O$_3$, the magnetic structure factor of the
(2,0,0)$_{\rm c}$ reflection is proportional to
\begin{equation}\label{StructureFactor}
  F(2,0,0)_{\rm c}\sim|-8\:\mu_1 + 7.39\:\mu_2|
\end{equation}
where the two contributions come from the two inequivalent
neodymium sites with potentially inequivalent moments. Because of
the negative phase between the two contributions, in the case of
identical moments there is a 92\% cancelation between the
contributions from the two sites.  Since the atoms with C$_2$
symmetry are slightly displaced from the cell edge, the
cancelation is incomplete. This also holds for nuclear scattering,
in which case $\mu$ is replaced by the scattering length
$b_\text{Nd}$.  For the (2,0,2)$_{\rm c}$ reflection the structure
factor is proportional to
\begin{equation}
  F(2,0,2)_{\rm c}\sim|8\:\mu_1 - 8\:\mu_2|.
\end{equation}
Assuming cubic symmetry, the non-zero intensity apparent in Fig.
\ref{ChalkPiPi}b then implies that the magnetic moments on the two
Nd sites are, in fact, different for this particular field and
temperature. Furthermore, the existence of inequivalent moments
implies that the evolution of the magnetic moment with field is
also slightly different between the two sites. As a result,
non-trivial field dependence of the scattering intensity may be
observed.

\begin{figure}[tb]
\centering
\includegraphics{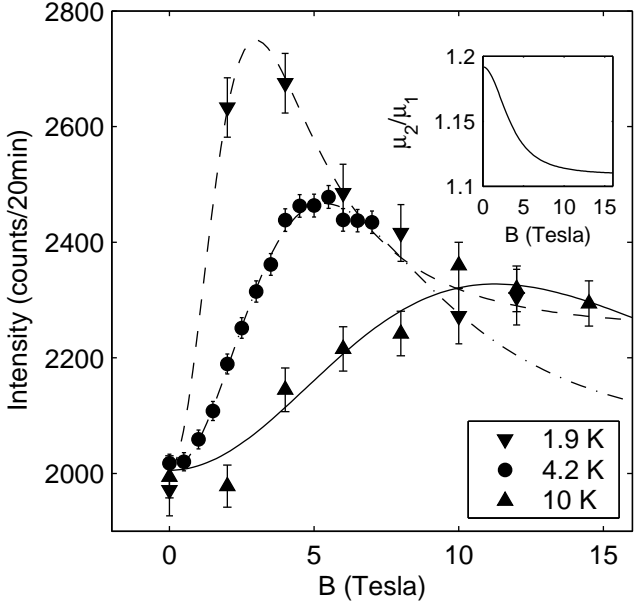}
\caption{\label{Brillouin} Field dependence of the scattering
observed at
  (1/2,1/2,0) at several temperatures for a reduced $x=0.18$ sample (NCCO18; see Table 1).
  The magnetic field was applied along [0,0,1]. The 1.9 K and 10 K
  data were taken at the Hahn-Meitner-Institut.  The 4.2 K data were taken
  at the NIST Center for Neutron Research and normalized to the former
  using additional data (not shown) for the temperature dependence at a
  constant field.  The lines are the result of a fit to a two-moment
  model, as described in the text. (inset) The ratio of the moment at the
  Nd(1) site to the moment at the Nd(2) site of cubic \NDO~ at 1.9 K, as extracted
  from the fit.  }
\end{figure}

In Fig. \ref{Brillouin}, we show the field dependence of the
scattering at (1/2,1/2,0) for the NCCO ($x=0.18$) sample at three
different temperatures. The lowest and highest temperature data
were taken using the two-axis diffractometer E4 at the
Hahn-Meitner-Institut and the 4.2 K data were taken using the BT2
spectrometer at the NIST Center for Neutron Research.  Temperature
scans taken at constant field (data not shown) were used to
cross-normalize the data sets taken at the two facilities.  Both
experiments employed a vertical-field magnet with field directed
along [0,0,1], but the magnetic field range of the 4.2 K data set
is limited by the 7 T maximum field of the NIST magnet.  We model
the behavior illustrated in Fig. \ref{Brillouin} by assuming that
the
moment on each Nd site evolves according to its own Brillouin
function\cite{Book:Kittel}
\begin{eqnarray}
\mu_{1,2}&=&g\mu_BJ_{1,2}B_{J_{1,2}}; \\
B_J&=&\frac{J+1/2}{J}\coth[(J+1/2)\eta] - \frac{1}{2J}\coth(\eta/2); \nonumber\\
\eta&=&\frac{g\mu_B B}{kT} \nonumber
\end{eqnarray}
and then combine the associated scattering intensity according to
the aforementioned structure factor in Eq. \ref{StructureFactor}.
The effect of the crystal field on the total angular momentum $J$
is unknown for \NDO, and we have fitted $J$ at each temperature
for both Nd sites. Once we fix the non-magnetic scattering to 2000
counts/20 min., the fit involves an overall
(temperature-independent) scale factor and two additional
parameters per temperature: $J_1$ and $J_2$. From the lowest to
highest temperature we find $J_1$=2.1(2), 2.4(1), 2.6(2), and
$J_2/J_1$=1.110(4), 1.093(4), 1.085(4). The Land\'{e} factor $g$
was fixed at the free ion value of 8/11 for Nd.  Clearly, the
two-moment model provides an excellent description of our data. In
the inset of Fig. \ref{Brillouin}, we plot the field evolution of
the ratio $\mu_2/\mu_1$ as extracted from the field dependence at
the lowest temperature. Our model indicates that the saturated
moments of the two sites differ by ~11\%. This is
consistent with the 13\% difference found for Er$_2$O$_3$ and the
77\% difference found for Yb$_2$O$_3$.\cite{Er2O3:Moon} From the
fits, we estimate the low-temperature moment of the Nd(1) atoms as
$gJ\mu_{\rm B} \approx 1.5\mu_{\rm B}$.  This is a reasonable
value, comparable to the experimentally observed value of
1.65$\mu_{\rm B}$ of cubic NdNi$_2$.\cite{skrabek63} The
two-moment model accounts not only for the complete field and
temperature dependence of the scattering intensity, but also
explains how what is essentially ferromagnetic scattering can
seemingly decrease with increasing magnetic field.

The results presented in Fig. \ref{Brillouin} indicate that \NDO~
is paramagnetic even at 1.9 K.  This is consistent with the fact
that Er$_2$O$_3$ ($T_{\rm N} = 3.4$ K) and Yb$_2$O$_3$ ($T_{\rm N}
= 2.3$ K) have relatively low N\'eel temperatures, and that the
\NDO~secondary phase has a lower effective dimensionality.  In a
separate measurement we did not observe spontaneous Nd ordering
(of \NDO~or NCCO) in our $x=0.10$ sample down to 1.4 K.

We note that a field-dependent peak was also observed at (1/2,0,0)
(equivalent to (1,1,0)$_{\rm c}$) and at (1/4,1/4,1.1) (equivalent
to (1,0,1)$_{\rm c}$) in Fig. \ref{10NSC}. As discussed in Sec.
VI, reflections at these positions are forbidden by the $Ia3$
glide-plane symmetry and our observation of such peaks is further
evidence of the lowered symmetry of the strained environment
inside the NCCO matrix.  Since Eq. (\ref{StructureFactor}) for the
structure factor at (1/2,1/2,0) represents a near cancelation of
contributions, a small symmetry breaking can result in scattering
intensities at (1/2,0,0) that are comparable to those observed at
(1/2,1/2,0).

\begin{figure}[b]
\centering
\includegraphics{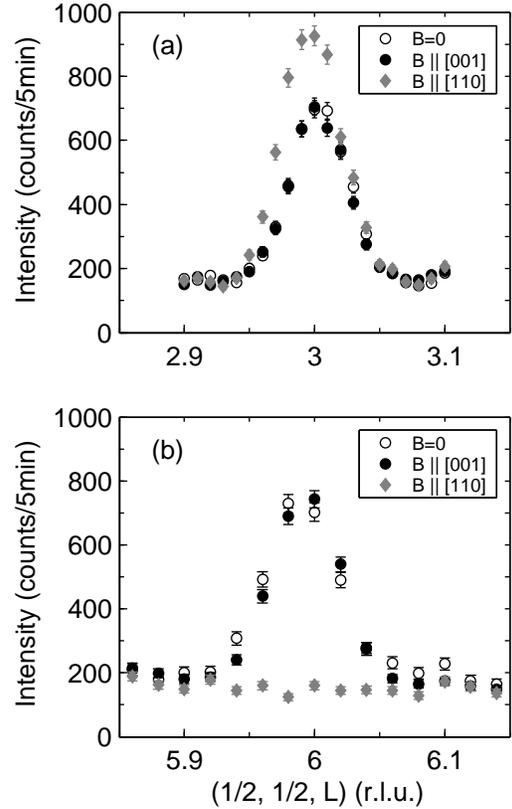}
\caption{\label{ChalkIntrinsic} Scans through NCCO magnetic peak
positions
  $(1/2,1/2,L)$ for a reduced $x=0.18$ sample (NCCO18; see Table 1).  The
  magnetic field was 2.6 T and the temperature was 5 K. The zero-field
  magnetic scattering only appears below 8 K.
  Whereas a
  magnetic field applied along the $c$-axis has no effect, an in-plane
  field along [1,1,0] slightly
  enhances the odd-integer-$L$ peak and completely suppresses the
  even-integer-$L$ scattering.  The data were taken at Chalk River
  Laboratories.}
\end{figure}

The field-effect results presented above establish that a magnetic
contribution from \NDO~ is present at (1/2,1/2,0), the
two-dimensional antiferromagnetic zone center of NCCO, and at
equivalent positions, but it leaves open the possibility that a
small fraction of the signal may be intrinsic to the copper-oxygen
sheets. One way to separate the two possible contributions is to
perform scans at reciprocal space positions with non-zero, integer
$L$, with a magnetic field along [0,0,1], such that the scattering
from NCCO and from \NDO~ is clearly separated. Experimentally,
this requires that the sample be aligned in the $(H,H,L)$
scattering zone and that a horizontal-field magnet be employed.
Only a few such magnets compatible with use at a neutron
diffractometer exist, and their maximum field is less than that of
vertical-field magnets.

Figure \ref{ChalkIntrinsic} shows the result of such a measurement
for a reduced $x=0.18$ sample, performed at Chalk River
Laboratories at two antiferromagnetic Bragg peak positions
corresponding to $(1/2,1/2,L)$ with both even and odd integer
values of $L$.  As-grown NCCO exhibits long-range magnetic order
of the copper moments even at high cerium concentrations, with
N\'eel temperature $T_{\rm N} \sim 80 - 100$ K for $x=0.16-0.18$.
\cite{mang03b} Along $(1/2,1/2,L)$, magnetic Bragg scattering is
observed at non-zero integer values of $L$.  The zero-field
magnetic intensity at these positions in reduced, superconducting
samples is relatively weak when compared to as-grown NCCO,
\cite{matsuda92,uefuji02,mang03b} and the antiferromagnetic volume
fraction decreases rapidly with increasing $x$ in the
superconducting phase.\cite{uefuji01} Moreover, it has been found
that the antiferromagnetic regions have a finite extent of 50-100
\AA~ in superconducting samples with $x\ge0.15$.\cite{uefuji02} In
our case ($x=0.18$), the scattering indeed is very weak, and only
appears below $\sim8$ K. From Fig. \ref{ChalkIntrinsic}, we see
that the application of a magnetic field along [0,0,1] has no
effect on the scattering intensity. However, a field parallel to
the copper-oxygen planes enhances the odd-$L$ peak intensity and
completely suppresses the even-$L$ peak. Presumably this is the
result of a non-collinear to collinear spin transformation similar
to what has previously been
reported,\cite{NCCO:SpinReorientation:Lynn} but in the present
case the orientation of the central spin (at the body-center
position; see Fig. 1) is rotated by $\pi/2$ resulting in the
elimination of even-$L$, rather than odd-$L$, peaks.

\section{Discussion}

\begin{figure}[!t]
\centering
\includegraphics[height=14.572cm]{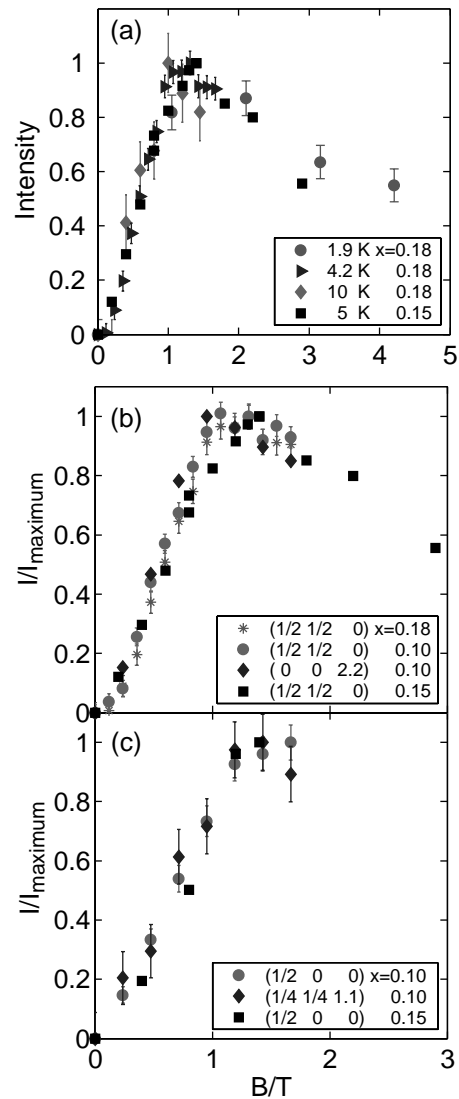}
\caption{\label{NatureFig} Field and temperature dependence of
magnetic
  scattering. (a) Arbitrarily scaled scattering intensity at (1/2,1/2,0)
  for a superconducting sample of NCCO (nominal cerium concentration
  $x=0.18$; \Tc = 20 K) at several temperatures, plotted as a function of
  B/T with the field along [0,0,1].  The results are compared with the
  data at $T = 5$ K of Kang \etal \cite{NCCO:SecondPhase:DaiNature}
  ($x=0.15$; \Tc=25 K). (b,c) Comparison of the results of Kang
  \etal~with data taken at $T=4$ K for a superconducting sample ($x=0.18$)
  and a non-superconducting sample ($x=0.10$). The magnetic field is
  applied along [1,$\bar{1}$,0] for (0,0,2.2) and (1/4,1/4,1.1) and along
  [0,0,1] in all other cases. Data were normalized by maximum
  intensity. Our samples are listed as NCCO10 and NCCO18 in Table 1.
  This figure is reproduced from Ref. \onlinecite{NCCO:SecondPhase:NatureComment} }
\end{figure}

Kang and co-workers reported magnetic-field-induced scattering at
(1/2,1/2,0), (1/2,0,0) and related reflections, and interpreted
their data as indicative of a field-induced quantum phase
transition from a superconducting to an antiferromagnetic state.
\cite{NCCO:SecondPhase:DaiNature} In a previously published
version of Fig. \ref{NatureFig},
\cite{NCCO:SecondPhase:NatureComment} we demonstrated that a
number of data sets taken at different temperatures, magnetic
fields, reciprocal space positions, and on different samples (both
superconducting and non-superconducting) could be scaled onto an
approximately universal curve of scattering intensity versus B/T.
We argued that these observations were inconsistent with a quantum
phase transition from a superconducting to an antiferromagnetic
state, and instead suggested that the data could be explained by a
two-moment paramagnetic model for \NDO, introduced in detail in
Sec. VIII. Kang and co-workers now acknowledge the presence of
\NDO~in their reduced samples and that this secondary phase does
exhibit a field-induced paramagnetic response which leads to an
enhancement of the scattering intensity at the reciprocal space
positions they
measured.\cite{NCCO:SecondPhase:DaiNatureComment,NCCO:SecondPhase:DaiPRB}

\begin{figure}[t]
\centering
\includegraphics{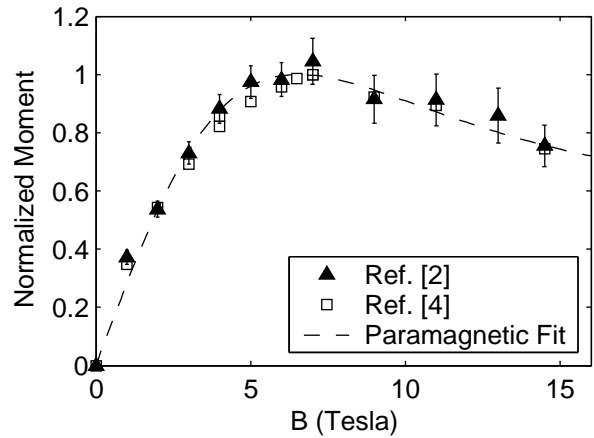}
\caption{\label{ZhangCompare} Experimental results ($x=0.15$;
$T=5$ K) for
  the field dependence of the normalized moment from
  Ref. \onlinecite{NCCO:SecondPhase:DaiNature}
  with \NDO~two-moment
  fit (dashed line), as described in the text. Also shown are the difference data from
  Ref. \onlinecite{NCCO:SecondPhase:Zhang}.
  }
\end{figure}

However, the authors of Refs. \onlinecite{NCCO:SecondPhase:DaiPRB}
and \onlinecite{NCCO:SecondPhase:DaiNatureComment} insist that
they have evidence for a NCCO contribution that can be effectively
extracted by, for example, subtracting the intensity at
(1/2,1/2,0) ((2,0,0)$_{\rm c}$) measured with ${\bf B}$ $||$
[1,-1,0] ([0,1,0]$_{\rm c}$) from that measured with ${\bf B}$
$||$ [0,0,1] ([0,0,1]$_{\rm c}$).  In both geometries the
scattering wavevector {\bf Q} is perpendicular to {\bf B} and,
under the assumption of cubic $Ia3$ symmetry and three-dimensional
long-range order, the contribution from \NDO~should be the same.
This procedure was attempted in Ref.
\onlinecite{NCCO:SecondPhase:Zhang}. In Fig. \ref{ZhangCompare},
we reproduce the difference data reported in that reference.  We
note that what is shown is the (normalized) moment, which is
proportional to the square root of the magnetic neutron scattering
intensity. \cite{NCCO:SecondPhase:Zhang} In addition, we reproduce
the original, uncorrected data of Kang \etal
\cite{NCCO:SecondPhase:DaiNature} taken with a magnetic field
along [0,0,1]. We note that the subtraction procedure does not
significantly alter the field dependence of the data. We also show
a fit of the uncorrected data to the two-moment paramagnetic model
of Sec. VIII, with fit parameters J$_1$=2.6(4) and
J$_2/$J$_1$=1.11(1), which describe the data just as well as the
theoretical treatment of Ref. \onlinecite{NCCO:SecondPhase:Zhang}.
We emphasize that our fit of the normalized data contains only two
fit parameters. Given that a quantum phase transition in NCCO and
paramagnetism of \NDO~ are fundamentally different physical
phenomena, we consider it incredible that the two should have the
same magnetic field dependence, and hence conclude that the
subtraction procedure is an unreliable means of correcting the
data. This may be in part due to a thin-film-like magnetic
anisotropy present in the secondary phase, or to experimental
difficulties in accurately normalizing the signal between the two
geometries.

Kang \etal
\cite{NCCO:SecondPhase:DaiNature,NCCO:SecondPhase:DaiNatureComment}
and Matsuura \etal \cite{NCCO:SecondPhase:DaiPRB} argue that the
inclusions of \NDO~in their samples are a bulk phase, and hence
physically distinct from the epitaxial, quasi-two-dimensional
nature we report here. Matsuura et al. conclude that \NDO~has
sharp Bragg peaks indicative of three-dimensional long-range order
on the basis of survey scans along $(1/2,0,L)$ and $(3/2,3/2,L)$
which consist of no more than one datum in each peak (Fig. 7 of
Ref. \onlinecite{NCCO:SecondPhase:DaiPRB}). However, close
examination of the (1/2,1/2,3) NCCO peak and the (1/2,1/2,4.4)
((2,0,4)$_{\rm c}$) \NDO~peak (shown in Figs. 8b and 8c of Ref.
\onlinecite{NCCO:SecondPhase:DaiPRB}) reveals that, whereas the
NCCO peak does seem to be resolution limited with a HWHM of $\sim
0.015$ r.l.u., the \NDO~peak is $\sim 2.5$ times wider, with a
HWHM of $\sim 0.04$ r.l.u. From parameters published in Ref. 3, we
estimate the momentum resolution at (1/2,1/2,4.4) to be only about
20\% broader than that at (1/2,1/2,3). Consequently, the large
width corresponds to a finite out-of-plane correlation length of
approximately 50 \AA, or about $5a_{\rm c}$. This is consistent
with our own data, and with the broad structural peaks reported by
Kurahashi \etal,\cite{NCCO:Superstructure:Kurahashi} but
incompatible with claims of three-dimensional long-range order. It
is furthermore consistent with the statement in the caption to
Fig. 4 of Ref. \onlinecite{NCCO:SecondPhase:DaiNature}, where the
additional reflections are characterized as very broad. We note
that the aluminium powder peak visible in Fig. 7 of Ref.
\onlinecite{NCCO:SecondPhase:DaiPRB} is a poor indicator of the
experimental resolution because the scan is not purely radial, and
hence contains a broadening component from the aluminium powder
ring. From the available data we conclude that the nature of the
\NDO~inclusions present in the samples studied in Refs.
\onlinecite{NCCO:SecondPhase:DaiNature, NCCO:SecondPhase:DaiPRB,
NCCO:SecondPhase:DaiNatureComment} are entirely consistent with
what we report here, i.e., an epitaxial structure, with an average
thickness of 50-100 \AA. Finally, we note that Kang et al. also
report magnetic-field induced scattering at (1/2,0,0), which
indicates that the glide-plane symmetry in their sample has also
been broken and, hence, as explained in Sec. VI, the structural
symmetery is lower than cubic.

The only observation of Refs. 2 and 3 that differs from our own
results is the field dependence of the (1/2,1/2,3) peak. As noted
at the end of Sec. IX, the volume fraction of the NCCO
antiferromagnetic phase strongly decreases with increasing cerium
concentration in reduced samples. We note that the superconducting
and antiferromagnetic volume fractions have been found to be
highly anticorrelated. \cite{uefuji01} Given the sample
inhomogeneity issues discussed in Sec. IV, it appears likely that
this (zero-field) magnetic response is a non-superconducting
remnant of the as-grown state, with modified oxygen concentration
and disorder.  The response to a magnetic field of this remnant
phase is a priori unclear, since the phase is altered by the
oxygen treatment from the as-grown state.  Our 5 K data taken in
the horizontal-field geometry and presented in Fig. 15 do not
indicate any change in the scattering at the (1/2,1/2,3) and
(1/2,1/2,6) reflections of NCCO for magnetic fields up to 2.6 T
oriented along [0,0,1].  Given that a field of 2.6 T constitutes a
significant fraction ($\sim 40\%$) of the upper critical field in
this geometry, \cite{NCCO:SecondPhase:DaiNature} the absence of a
field effect suggests that the scattering at the reflections of
the NCCO antiferromagnetic minority phase is unrelated to the
suppression of superconductivity in the NCCO majority phase.

In Refs. \onlinecite{NCCO:SecondPhase:DaiNature} and
\onlinecite{NCCO:SecondPhase:DaiPRB} a small enhancement of the
(1/2,1/2,3) intensity was reported in fields up to 4 T along
[0,0,1], also at 5 K, for a $x=0.15$ sample.  The cerium
concentrations of the two samples are somewhat different ($x=0.18$
versus $x=0.15$) and the effect of a magnetic field on the
antiferromagnetic NCCO minority phase may have a subtle doping
dependence.  Alternatively, since the Nd moment increases quite
dramatically in a $1/T$ fashion, and because it contributes to the
NCCO structure factor,\cite{matsuda92} small temperature
differences between two scans could, in principle, appear as a
change in intensity.  A third possible origin for the observed
field effect reported in Refs.
\onlinecite{NCCO:SecondPhase:DaiNature} and
\onlinecite{NCCO:SecondPhase:DaiPRB} could be due to the torque
exerted on the sample for field configurations ${\bf B} \parallel
[0,0,1]$. If the mount used in the experiment is not sufficiently
rigid the sample may rotate, resulting in a difference in
intensity between two scans. This is especially serious
considering our result reported in Fig. \ref{ChalkIntrinsic} in
which a field directed along the [1,1,0] direction increases the
amount of scattering at the (1/2,1/2,3) reflection. Small changes
in the alignment of the sample can either increase or decrease the
observed scattering. Finally, we note that the (1/2,1/2,3) data in
Fig 9d of Ref. \onlinecite{NCCO:SecondPhase:DaiPRB} have
relatively large error bars and are not entirely inconsistent with
the absence of a field dependence.

An earlier study of \NCCO~($x=0.14$) in [0,0,1] fields up to 10 T
failed to detect any enhancement of the scattering intensity at
the (1/2,3/2,0) reflection.\cite{NCCO:SecondPhase:YamadaNull} The
lowest temperature at which data were taken in that experiment was
a relatively high 15 K. Because of the strong enhancement of the
Nd moment at low temperatures, this would have made detection of
the effect from the \NDO~phase more difficult. Also, measurements
were only conducted at the (1/2,3/2,0) position. This is an
allowed NCCO magnetic reflection, unlike (1/2,1/2,0), which is
disallowed by the spin structure.  Consequently, the additional
contribution from \NDO~would also have been difficult to detect
due to the prominence of the large signal from the NCCO
antiferromagnetic phase regions of the sample, since the volume
fraction of the NCCO antiferromagnetic phase is still relatively
large for $x=0.14$.\cite{uefuji01} A subsequent experiment,
conducted on the related electron-doped compound
Pr$_{1-x}$LaCe$_x$CuO$_4$ (PLCCO, $x=0.15$, $T_{\rm c} = 16$ K),
failed to detect any magnetic field effects at the (3/2,1/2,0)
reflection in fields of up to 8.5 T along
[0,0,1].\cite{NCCO:SecondPhase:PLCCO} This is consistent with our
observations. PLCCO has the advantage that La is a non-magnetic
ion, and the Pr moment is an order of magnitude weaker than that
of Nd at low temperatures. Therefore, any complicating magnetic
effects from a (Pr,La,Ce)$_2$O$_3$ secondary phase should be
considerably weaker. The failure to observe a magnetic field
effect deep inside the superconducting dome at $x=0.15$ is
consistent with our statements that such effects in NCCO can be
entirely accounted for by the \NDO~secondary phase. We note that
some effects qualitatively different from those in NCCO were
reported for an $x=0.11$ sample that lies on the boundary of the
antiferromagnetic and superconducting regions of the
Pr$_{1-x}$LaCe$_x$CuO$_4$ phase
diagram.\cite{NCCO:SecondPhase:PLCCO} It appears that those are
attributable to a disturbance of the antiferromagnetic
Pr$_{1-x}$LaCe$_x$CuO$_4$ minority phase and unconnected with
superconductivity.

\section{Summary}

In summary, we have observed that when \NCCO~is exposed to a
reducing environment a small fraction of the crystal decomposes.
One of the decomposition products is \NDO, which exists
epitaxially in a strained cubic bixbyite structure. The structure
is long-range correlated parallel to the copper-oxygen planes, but
only short-range ordered perpendicular to the planes.  Application
of a magnetic field polarizes the Nd atoms, leading to an
enhancement of the magnetic neutron scattering intensity at
positions coincident with \NDO~structural peaks. A simple
two-moment model for the \NDO~paramagnetism gives an excellent
description of our data. The extensive data presented here are
inconsistent with the notion
\cite{NCCO:SecondPhase:DaiNature,NCCO:SecondPhase:DaiNatureComment,NCCO:SecondPhase:DaiPRB,
NCCO:SecondPhase:Zhang} of a field-induced quantum phase
transition from a superconducting to an antiferromagnetic state of
\NCCO.  We note that the presence of the secondary phase should
also be taken into account in the analysis of other experiments on
\NCCO, such as transport measurements.

\acknowledgments

We would like to thank N.  Kaneko for his efforts in constructing
the reduction furnace and maintaining the crystal growth facility
at Stanford University, and A. Arvanitaki for assistance in
orienting the TEM sample. We are also grateful to J.M Tranquada,
H. Eisaki, and N.P. Armitage for valuable discussions.  We would
like to thank J.W. Lynn for technical assistance with the neutron
scattering measurements at NIST. Finally, we wish to acknowledge
P. Dai, H.J. Kang, J.W. Lynn, M. Matsuura, and S.C. Zhang for
discussing their data with us.  SSRL is supported by the DOE
Office of Basic Energy Sciences, Division of Chemical Sciences and
Material Sciences. The work at Stanford was furthermore supported
by the US Department of Energy under Contracts No.
DE-FG03-99ER45773 and No. DE-AC03-76SF00515, and by NSF CAREER
Award No. DMR9985067.

\newpage

\bibliography{MangPRB2}

\begin{thebibliography}{35}
\bibitem[{Tem()}]{TemperatureNote}
\bibinfo{note}{In previous publications we reported the furnace control
  temperature instead of the sample temperature.}
\bibitem[{VF()}]{VF}
\bibinfo{note}{In our original publication
  \cite{NCCO:SecondPhase:NatureComment} we gave an estimate of 0.01-0.1\% for
  the typical secondary phase volume fraction. Table I contains refined
  estimates, ranging two orders in magnitude, from $\sim 0.02$\% to $\sim 2$\%,
  depending on the reduction conditions.}
\end{thebibliography}

\end{document}